\documentstyle{mn}

\begin{document}

\title[Arp 220 Super Star Clusters]
{Photometric Properties of the Arp 220 Super Star Clusters}

\author[Y. Shioya et al.]
{Yasuhiro Shioya$^1$\thanks{shioya@astr.tohoku.ac.jp}, 
Yoshiaki Taniguchi$^1$, \& Neil Trentham$^2$\\
$^1$Astronomical Institute, Graduate School of Science, 
Tohoku University, Aramaki, Aoba, Sendai 980-8578, Japan\\
$^2$Institute of Astronomy, University of Cambridge, Madingley Road, Cambridge
       CB3 0HA, UK}

\maketitle
\begin{abstract}
We investigate the photometric properties of six super stellar 
clusters (SSCs) seen in both the optical and 
near-infrared Hubble Space Telescope images of the local 
ultraluminous starburst galaxy Arp 220.
Three of the SSCs are located in the central 0.5 kpc region.  
The remaining three are in the circumnuclear region between 
0.5 kpc and 2.5 kpc from the centre. 
Comparing the observed 
spectral energy distributions (SEDs) of the SSCs 
with the Starburst99 models of Leitherer et al., 
we confirm that all the three nuclear SSCs 
are heavily obscured ($A_V \sim 10$ mag). 
Considering the results from this comparison
in conjunction with measurements of
the near-infrared CO absorption index and 
of millimetre CO line widths and luminosities,
we estimate the ages of the nuclear SSCs to be $10^7 - 10^8$ yrs.  
The bolometric luminosity of the three nuclear SSCs 
is at most one-fifth of the total bolometric 
luminosity of Arp 220. 
On the other hand, the circumnuclear SSCs 
have little internal extinction ($A_V \leq$ 1 mag).  These contribute
negligibly to the total bolometric luminosity.   
\end{abstract}

%------------------------------------------------------------------------------------
%------------------------------------------------------------------------------------

\section{INTRODUCTION}

Ultraluminous infrared galaxies (ULIGs) are the most luminous objects
in the local universe (their bolometric luminosities 
$ > 10^{12} {\rm L}_{\odot}$). 
The ULIGs are undergoing major dissipative collapses, which are 
probably triggered by mergers 
(Clements et al.~1996; Murphy et al.~1996). 

Whether the ULIGs are powered by starbursts or active galactic 
nuclei (AGN) has remained unknown since their discovery 
because of huge amounts of internal extinction 
along our lines of sight to their centres  
(for a review see Sanders \& Mirabel 1996). 
However, recent 
mid-infrared spectroscopic work (Genzel et al.~1998) 
suggests that the major energy sources of most local
ULIGs are nuclear starbursts. 
This provides an important motivation for studying 
the star formation in the centres of ULIGs in detail. 

The nearest and best-studied ULIG is the star-forming
(Genzel et al.~1998) galaxy Arp 220 
(far-infrared luminosity $1.5 \times 10^{12} \, {\rm L}_{\odot}$)
at a distance of 74 Mpc 
(assuming $H_0$ = 75 km s$^{-1}$ Mpc$^{-1}$ and 
$\Omega_{0}=1$; de Vaucouleurs et al.~1991).     
Detailed imaging of the centre of Arp 220 with the
{\it Hubble Space Telescope} has revealed a number of
super star clusters (SSCs; 
Shaya et al.~1994, Scoville et al.~1998). 
These nuclear SSCs appear to be a generic feature of luminous
merging galaxies (Lutz 1991; Ashman \& Zepf 1992; 
Holtzman et al.~1992; Zepf \& Ashman 1993; Surace et al.~1998; 
Surace \& Sanders 1999). 
Surace et al. (1998) and Surace \& Sanders (1999) 
evaluated the luminosities of SSCs in warm ULIGs and concluded that 
the combined contributions of the all the individual detected
circumnuclear SSCs 
to the bolometric luminosities are small.
They also showed that for some warm ULIGs 
the de-reddened luminosities of putative nuclei are not able to 
account for the bolometric luminosity and that a large fraction of 
the bolometric luminosity must arise from sources 
undetected at both optical and near-infrared wavelengths. 

In this paper, we compare the observed optical and near-infrared 
spectral energy distributions (SEDs) of the Arp 220 SSCs 
with the Starburst99 model SEDs of Leitherer et al.~(1999) and 
estimate their ages, masses, and luminosities, 
along with the internal extinction along our lines of sight to them. 
This is the first attempt to analyse by SED fitting methods
the properties 
of SSCs in the centre of Arp 220, 
which is colder (Sanders et al.~1988) than the ULIGs studied
by Surace et al.~(1998).  
These results will let us evaluate how much of 
the very substantial amount of star formation currently happening 
in Arp 220 is in the SSCs, at least in the ones which are not 
completely invisible at optical and near-infrared wavelengths 
due to internal extinction. 
Recently, Soifer et al. (1999) presented the images of 
Arp 220 from 3.45 to 24.5 $\mu$m. 
Since Genzel et al. (1998) derived the value of $A_V=45$ mag 
based on the mid-infrared hydrogen recombination lines 
(Br$\beta$ $\lambda$ 2.17 $\mu$m, Br$\alpha$ $\lambda$ 4.05$\mu$m 
and Pf$\alpha$ $\lambda$ 7.46 $\mu$m), 
the mid-infrared sources observed by Soifer et al. (1999) 
must be highly obscured objects. 
This suggests that what Surace et al.~(1998) found to be true
in the warm ULIGs, specifically that the contribution of the
observed clusters to the bolometric luminosity is small, is 
also true in Arp 220.  We now investigate this in detail by
studying the energy outputs of the SSCs themselves.

%------------------------------------------------------------------------------

\section{METHODS}

\subsection{Identifications of SSCs}

The {\it Hubble Space Telescope} images of the core of Arp 220 
show eleven SSCs at optical ($V$-, $R$-, and $I$-band) wavelengths 
(Shaya et al.~1994) 
and twelve SSCs at near-infrared ($J$-, $H$-, and $K$-band) 
wavelengths (Scoville et al.~1998). 
In this paper, we call the F110W filter (1.1 $\mu$m) as $J$ filter, 
though the standard $J$ filter is at 1.25$\mu$m. 
We combine these datasets in order to obtain a set of SSCs that are 
detected at all wavelengths. 
This allows us to sample the SEDs over as wide 
a range in wavelength as possible. 

Three SSCs are located in the outer regions of the core 
-- we expect dust extinction to be smallest here, 
so that these SSCs should be seen at all wavelengths. 
Given the published coordinates, there is a slight offset 
between the near-infrared and optical positions of these SSCs 
(see the left panel of Figure 1). 
However, if we rotate the near-infrared images by $-4^{\circ}$ 
around the nuclear SSC associated with the western nucleus, 
the positions of the star clusters in the two images 
are almost coincident (see the right panel of Figure 1). 
Given the probable low extinction along these lines of sight, 
we regard this astrometrical solution as likely to be the 
correct one. 
In addition, given this astrometry, 
we then find that three nuclear SSCs (hereafter N1, N2, and N3) 
are coincident in the optical and near-infrared images, 
in addition to the three circumnuclear ones 
(hereafter C1, C2, and C3).

\subsection{Spectral energy distributions of the SSCs}

In Figure 2, we show the observed SEDs of the six SSCs.
We use the photometric data 
published by Shaya et al.~(1994; $VRI$ bands) and 
by Scoville et al.~(1998; $JHK$ bands) for SSC N2 -- N3 and C1 -- C3. 
In the case of SSC N1, 
we have used $HST$ archival data to measure the optical fluxes using the
same 0.92 arcsec $\times$ 0.58 arcsec aperture used by Scoville 
et al.~(1998) for the near-infrared measurements (Shaya et al.~used
a smaller aperture in their analysis).  
The magnitudes of SSC N1 are 21.96 mag and 19.36 
for $R$-band (F702W) and $I$-band (F785LP) respectively. 
This SSC was not detected in the $V$-band (F555W). 

All three nuclear SSCs show a peak at 1.6 $\mu$m, whereas all three
circumnuclear SSCs have SEDs that rise towards bluer wavelengths.
This is a very important difference and is immediately suggestive
of far more dust extinction along the lines of sight to the nuclear
SSCs than along the lines of sight to the circumnuclear ones. 

We now compare the SEDs with the Starburst99 spectral synthesis 
models of Leitherer et al.~(1999). 
One of the advantages of the Starburst99 is 
that it tell us the evolution of the strength of CO index 
which is useful to constrain the range of ages. 
When fitting model SEDs to observed ones in the presence of 
internal extinction, there are several 
parameters that we need to consider: 
\vskip 1pt
\noindent
1) the mode of star formation.  For example, the star formation may 
occur continuously (the constant star formation, or CSF model), 
may occur in a short, almost instantaneous, burst (the instantaneous 
starburst, or ISB model), or may occur in a way that 
varies with time in a complex manner. 
For the blue circumnuclear SSCs, there is no evidence that star 
formation has been ongoing in these isolated 
clusters for any considerable time, and we adopt an 
ISB model for describing star formation in these systems. 
On the other hand, the nuclear star-forming regions of Arp 220 are 
probably at least 10$^{8}$ years old (Mouri \& Taniguchi 1992; 
Prestwich, Joseph, \& Wright 1994; Armus et al.~1995; Larkin et al.~1995). 
Yet there still appear to be ionizing sources in the nuclear region 
(Larkin et al.~1995, Kim et al.~1995, and Goldader et al.~1995). 
Therefore an ISB model will not work here, and we  
approximate star formation in the nuclear SSCs 
by a CSF model. 
On the other hand, there is a possibility that the nuclear SSCs 
are not the ongoing starburst discussed in the 
above papers but post-starburst 
and we also consider an ISB model for the nuclear SSCs, 
although we regard the CSF one as the more likely;  
\vskip 1pt
\noindent
2) the stellar IMF.  We adopt a Salpeter IMF with 
an upper mass cutoff of $M_{\rm u} = 100 \, {\rm M}_\odot$ and a lower mass
cutoff of $M_{\rm l} = 1 \, {\rm M}_\odot$. 
We note that $M_{\rm l}$ might be larger in a violent 
star-forming region (see e.g.~Goldader et al.~1997) and investigate
the effect of the IMF on our results in the next section; 
\vskip 1pt
\noindent
3) the initial gas metallicity. 
We assume solar metallicity ($Z=0.02$). 
Since the metallicity of galactic centre may be larger than 
the solar value, we also study the case of $Z=0.04$ (see the next section); 
\vskip 1pt
\noindent
4) the age at which we observe the star clusters.  
We leave this as a free parameter in the range of $2 \times 10^5$ yr 
to $1 \times 10^9$ yr;
\vskip 1pt
\noindent
5) the effect of extinction.  
The total extinction can be regarded as the sum of 
two parts: 
1) extinction from dust in Arp 220 along our line of sight to the SSCs, and 
2) extinction from dust {\it within} the SSCs. 
The relative importance of these two regimes can be tested as follows. 
A screen model may be used to describe extinction along our line of sight to
the SSCs, and 
the ``onion-skin" model of Surace and Sanders (1999) may be used to quantify
the extinction from dust within the SSCs. 
This comparison is made in Appendix A, where we show that (1) is the more
important regime.  We consequently adopt a screen model and 
leave the absolute value $A_V$ of 
the extinction in the $V$-band  as a free parameter 
to be fit to the data, 
but impose the constraint that the extinction 
must vary with wavelength according to the extinction curve of 
Cardelli et al.~(1989);
\vskip 1pt
\noindent
6) nebular emission. Starburst99 includes 
the nebular continuum emission which is proportional to 
the flux of ionizing photons. 
If all of the Lyman continuum photons are used to ionize the surrounding gas, 
the equivalent width of H$\alpha$ can reach 1000 \AA~and the $R$-band flux 
becomes twice as large as that from the continuum. 
However, the {\it measured} equivalent width of the H$\alpha$ emission line 
in Arp 220 is only  
20 -- 30 \AA (Armus et al.~1989, Veilleux et
al.~1995).
If the equivalent width of H$\alpha$ emission of each SSC is 
the same as the total ones, the contribution of emission lines 
to the broadband photometry is then negligibly small.  We neglect it here; 
\vskip 1pt
\noindent
7) dust emission. 
Hot dust emission was required to interpret a $K$-band flux excess 
in previous studies (Surace \& Sanders 1999 for warm ULIGs; 
Mazzarella et al. 1992 for the nuclei of Arp 220). 
However, there is little $K$-band flux excess for our sample. 
We ignore the hot dust emission in our models. 
\vskip 3pt
\noindent
Using maximum likelihood techniques, 
we determine the best-fitting model SEDs 
to the observed ones for the six SSCs.  
The likelihood is defined as 
\begin{equation}
L = \prod_i^6 \exp \left\{ - \frac{1}{2} \left[ 
\frac{F_{o,i}- a \cdot F_{m,i}}{\sigma_i}
\right]^2 \right\}
\end{equation}
where 
$F_{o,i}$, $F_{m,i}$ and $\sigma_i$ are 
the observed flux, the template flux, and the uncertainty of 
the observed flux of $i$-th band. 
The uncertainty of the photometry we adopted is $\pm 0.4$ mag 
for Shaya et al. (1994) and $\pm 0.1$ mag for Scoville et al. (1998). 
A scale factor $a$ is obtained as 
\begin{equation}
a = \frac{\displaystyle \sum_{i=1}^6 F_{m,i} F_{o,i}/\sigma_i^2}
{\displaystyle \sum_{i=1}^6 (F_{m,i})^2/\sigma_i^2}
\end{equation}
from the condition of $\partial L/\partial a=0$. 
The results are summarized in Table~1 and 
the SEDs of best-fitting models are presented in Figure 2.  

\section{RESULTS AND DISCUSSION}

\subsection{Properties of the nuclear SSCs}

Figure 3 is a summary of the region of 
plausible parameters in the $A_V$ - age plane for the nuclear SSCs. 
The results for the CSF model are shown in figure 3 (a) - (c) and 
those for the ISB model are shown in figure 3 (d) - (f). 

The hatched region is the region 
of plausible parameters at the 99 \% confidence level. 
We immediately see that the extinction $A_V$ along our lines of
sight to the nuclear SSCs is 
very large ($\sim 10$ mag). 
These values are slightly larger than those evaluated by 
Shaya et al.~(1994) while slightly smaller than those of 
Scoville et al.~(1998). 
These differences can be attributed to the extinction curves 
adopted in the analyses  
[Shaya et al.~use the extinction curve of 
Savage \& Mathis (1979), 
while Scoville et al.~use that of Rieke \& Lebofsky (1985)]. 

On the other hand, it is difficult to determine the age of SSCs, due to 
strong parameter coupling with the extinction.  A small change in
$A_V$ mimics a substantial change in age in the sense that both shift
the SED towards redder wavelengths.  

We explain the rough shapes of the relevant parameter space as follows.
If the stellar populations are very young (age $ < 10^7$ yr), 
$A_V$ must be very large, 
since the intrinsic slope of the continuum is very blue and a great deal of
extinction is required to make it as red as is observed. 
For stellar populations older than several $\times 10^7$ yr, 
the allowed values of $A_V$ do not vary strongly with age  
since the slope of continuum does not vary with the age of the stars. 
At the age of $\sim 10^7$ yr in the ISB model, 
the plausible value of $A_V$ becomes small temporarily due to the presence 
of a large number of red supergiants at exactly this age. 

Therefore, we cannot determine the ages of the nuclear SSCs this way. 
We must use additional measurements.  For N1, there exists an independent
constraint on the mass of the SSC from radio and millimetre observations.
The mass of N1 as determined by the SED fit represents a locus of
points in the $A_V$-age plane, and is shown in Figure 3 (the dashed line).  

The position of N1 corresponds to the 
western nucleus observed to exhibit radio continuum, 1.3 mm dust continuum 
and the CO (1-0) and (2-1) lines (Scoville et al. 1997; Downes \& Solomon 1998; 
Sakamoto et al. 1999). 
Millimetre spectroscopy (of CO lines)
can be used to determine both a dynamical mass 
(from the line widths) and a molecular gas mass (from the line luminosities). 
From the results of Downes \& Solomon (1998) and Sakamoto et al.~(1999), 
we estimate a dynamical mass of this nucleus of
$2 \times 10^9 {\rm M}_{\odot}$ 
and a gas mass of $1 \times 10^9 {\rm M}_{\odot}$ 
[adopting the
inclination parameter of Scoville et al.~(1998)]. 
The mass of the stellar component must 
be less than the difference between these two masses, which is  
$1 \times 10^9 {\rm M}_{\odot}$. 
If there are other (highly obscured) stellar components present within 
the region, 
this number is an upper limit; if all the mass of N1 is in gas and observed
stars, then this is an equality.  
The dashed lines in figure~3 (a) and (d) show 
the locus of SSCs with the mass of $1 \times 10^9{\rm M}_{\odot}$. 
The region of plausible parameters is now restricted to that to the  
left of the lines. 
The upper limits on the age of N1 are now $2.0 \times 10^8$ yr 
for the CSF model and $5.0 \times 10^7$ yr 
for the ISB model. 
The reason why we can set an upper limit of age is that 
the luminosity-to-mass ratio of the stellar cluster 
becomes smaller as it gets older and 
consequently a more massive cluster is required 
to radiate the observed flux. 

The dynamical masses of N2 and N3 have not been measured. 
Although the eastern nucleus of Arp 220 has been detected in CO, 
the position of it corresponds to neither N2 nor N3. 
Instead it is an entirely different object (hereafter we call it as N4),
coincident with the nuclear source SE in the near-infrared images of
Scoville et al.~(1998).  
Based on the fact that N1 and N4 only exhibit
non-thermal radio continuum, 
thermal dust continuum, and molecular gas,  
Scoville et al.~(1998) suggested that these two objects represent the
most substantial mass concentrations in the centre of Arp 220. 

We now consider yet another way to constrain the range of ages of the
nuclear SSCs. 
The strength of CO absorption index at 2.29 $\mu$m constrains  
the ages since it is sensitive 
primarily to the presence of red K supergiants, which are
stars with a narrow age distribution, 
particularly in the ISB models. 
The CO index of the centre of Arp 220 is known to be strong: 
$0.24 \pm 0.01$ mag (Goldader et al.~1995), 
$0.20 \pm 0.05$ mag (Shier et al.~1996), 
and $>0.20$ mag 
[Armus et al.~(1995) for the central $0.7 \times 2$ arcsec$^2$]. 
Although the CO indices of each SSC are not individually observed, 
if we assume that all three nuclear SSCs have a CO index 
larger than 0.20 mag, then we get  
following additional constraint on the ages:  
$8.1 \times 10^6 \le {\rm age/yr} \le 4.5 \times 10^7$ 
for the ISB model, or 
$1.2 \times 10^7 \le {\rm age/yr} < 1.2 \times 10^8$ 
for the CSF model. 
In the case of the ISB model, 
the CO index is 0 mag until the age of $7.1 \times 10^6$ yr and 
become strong rapidly at later times,
reaching 0.2 mag at the age of $8.0 \times 10^6$ yr,  
which corresponds to the main-sequence lifetime of the stars with mass of 
25 ${\rm M}_{\odot}$, the main progenitors of the K supergiants. 
After that, the CO index decreases gradually and 
becomes smaller than 0.2 mag 
at $4.5 \times 10^7$ yr. 
In the case of the CSF model, 
since newly formed red supergiants keep the CO index strong, 
the CO index continues to be larger than 0.2 mag for a longer time than 
for the ISB model. 
The strength of the CO index is modified slightly by dust as follows. 
Since the continuum level is determined  
on the blue side of CO band, dust extinction has the effect of causing
us to underestimate the continuum and therefore to underestimate the CO
index.   Ignoring dust emission at 2.3 microns has the same effect since
the SED of the dust component is rising towards longer wavelengths.
If the intrinsic value of CO index is larger than we think,  
then the upper limit of the age of SSCs become smaller.  Ignoring
the effects of dust therefore does not invalidate the upper limits we
present.  
 
Rieke et al. (1985) reported that the equivalent width of H$\beta$ 
absorption within the central 2.5 arcsec of Arp 220 is 4 \AA. 
This value of the Balmer absorption is incompatible with such a strong 
CO index for single stellar cluster. 
Probably this discrepancy arises because the Balmer feature originates from
old foreground (possibly progenitor galaxy) stars that happen to lie at low
optical depth and are quite unrelated to the SSCs.
The flux of the nuclear region of Arp 220 at very blue wavelengths is
therefore likely to be dominated by foreground stars at low skin-depth.  
Recall that the nuclear SSCs were not observed in the $V$-band so the
extinction at these short wavelengths is probably quite high.  

We can now estimate the maximum bolometric luminosity
emitted from the nuclear star clusters from the above results. 
For the ISB model, the maximum bolometric luminosities  
of N1, N2, N3 are $1.0 \times 10^{11}{\rm L}_{\odot}$, 
$1.2 \times 10^{10}{\rm L}_{\odot}$ and $4.1 \times 10^{10}{\rm L}_{\odot}$. 
The sum of these luminosities is $1.5 \times 10^{11}{\rm L}_{\odot}$ 
which is about one-tenth of the bolometric luminosity of Arp 220 
($1.5 \times 10^{12}{\rm L}_{\odot}$, Sanders et al. 1988). 
For the CSF model, the maximum bolometric luminosities  
of N1, N2, N3 are $1.9 \times 10^{11}{\rm L}_{\odot}$, 
$2.2 \times 10^{10}{\rm L}_{\odot}$ and $7.6 \times 10^{10}{\rm L}_{\odot}$. 
The sum of these is $2.9 \times 10^{11}{\rm L}_{\odot}$ 
which is about one-fifth of the bolometric luminosity of Arp 220. 
We can therefore state with some confidence that
if most of the total bolometric luminosity of Arp 220 
arises from the nucleus, more than four-fifths of it arises 
from highly obscured starbursts (or possibly an AGN). 

The results depend on the IMF as follows. 
A steeper IMF decreases the implied ages of the star clusters.
For example, IMF with ($\alpha=3.30$, $M_{\rm u}=100{\rm M}_{\odot}$, 
$M_{\rm l}=1{\rm M}_{\odot}$) 
decreases 
the maximum ages derived from the stellar mass ($<10^9{\rm M}_{\odot}$) to  
$1.6 \times 10^7$ yr for the CSF model or $1.4 \times 10^7$ yr for 
the ISB model. 
On the other hand, a truncated IMF increases the ages.
For example, an IMF with ($\alpha=2.35$, $M_{\rm u}=30{\rm M}_{\odot}$, 
and $M_{\rm l}=1{\rm M}_{\odot}$) increases the upper limits on the
ages to $2.4 \times 10^8$ yr 
for the CSF model and $5.2 \times 10^7$ yr for the ISB model. 
The evolution of CO index also depends on the IMF, 
since the number of red supergiants per unit total stellar mass 
at any time is a function of the IMF.
If we adopt the steeper IMF as above, 
the range of ages when the CO index is larger 
than 0.2 mag is $1.1 \times 10^7 \le {\rm age/yr} \le 7.1 \times 10^7$ 
for the CSF model or $8.2 \times 10^6 \le {\rm age/yr} \le 3.6 \times 10^7$ 
for ISB model. 
For the truncated IMF, it is  
$1.0 \times 10^7 \le {\rm age/yr} \le 1.6 \times 10^8$ for the CSF model or 
$8.1 \times 10^6 \le {\rm age/yr} \le 4.5 \times 10^7$ for the ISB model. 
The bolometric luminosity emitted from the nuclear 
star clusters also depends on the IMF, since we have imposed a constraint
on the mass.  
The bolometric luminosity ranges between $2.9 \times 10^{11}{\rm L}_{\odot}$ 
(CSF model with standard IMF) and $1.4 \times 10^{11}{\rm L}_{\odot}$ 
(ISB model with steeper IMF);  the basic conclusions are unchanged. 

The metallicities of many luminous
elliptical galaxies (the probable fate of
ultraluminous galaxies like Arp 220) are often considerably higher
than solar (Faber 1973).
It is therefore important to investigate
how the results
change if the metallicity is increased to twice solar (section 3.1, p.10),
comparable to the values in the most luminous ellipticals (Faber 1973).
For the SED fitting, the larger metallicity causes only a very
small change,
but the length of time when the CO index is strong
is considerably longer.
The range of ages when the CO index is larger than 0.2 mag is
$8.5 \times 10^6 \le {\rm age/yr} \le 3.0 \times 10^8$ for the CSF model or
$6.9 \times 10^6 \le {\rm age/yr} \le 5.4 \times 10^7$ for the ISB model.
The maximum bolometric luminosities are
$3.4 \times 10^{11}{\rm L}_{\odot}$ for CSF model and
$1.0 \times 10^{11}{\rm L}_{\odot}$ for ISB model;
the basic conclusions are also unchanged.

\vspace{1pc}
It is also useful to estimate how much of the difference between the
total bolometric luminosities and the luminosities of the observed SSCs
can be made up if one includes the contribution from N4, the SE nucleus
seen in the near-infrared but not the optical images. From SED fitting, 
we derive a value of $A_V \sim 12$ mag (Figure~4), 
much smaller than the value of $A_V \sim 20$ mag evaluated by 
Scoville et al.~(1998).  
This discrepancy comes from the fact that we use 
all three of the $J$-, $H$-, and $K$- band fluxes for SED fitting, 
whereas Scoville et al.~use only the $H-K$ colour. 

If, as for N1, we assume that the maximal mass of N4 is the dynamical
mass minus the gas mass, 
and assume that the strength of the CO index
is greater than 0.2 mag, we derive a maximum intrinsic luminosity of
N4 of $2.0 \times 10^{10}{\rm L}_{\odot}$ for ISB model or  
$3.7 \times 10^{10}{\rm L}_{\odot}$ for CSF model.
 
Adding these values to the summation of the bolometric luminosities  
of N1 - N3, we derive a maximum bolometric luminosity 
emitted from the nuclear SSCs of 
$1.73 \times 10^{11}{\rm L}_{\odot}$ for the ISB model and 
$3.22 \times 10^{11}{\rm L}_{\odot}$ for the CSF model. 
These values are still much smaller than the total bolometric 
luminosity of Arp 220. 
Again, this result is essentially not affected by the IMF or metallicity. 

Therefore the conclusion that most of the bolometric luminosity  
we see originating from the nucleus of Arp 220 comes from heavily 
obscured star formation seems unavoidable.  
At optical and near-infrared wavelengths we see only a small 
fraction (less than one-fifth) of it. 
The fact that the total luminosity of observed sources 
cannot account for the bolometric luminosity 
is similar to those derived for the putative nuclei 
of warm ULIGs (Surace \& Sanders 1999). 

\subsection{Properties of the circumnuclear SSCs}

Figures~3 (g) - (i) summarize the allowed regions 
on the $A_V$ - age plane for the circumnuclear SSCs. 
The data require total extinctions of $A_V < 1$ mag, far less  
than for the nuclear SSCs.  Because of this low extinction, the cluster
ages, masses, and luminosities are much better constrained by the SED fits
than for the nuclear clusters. 
We find that the upper limits on the ages of the SSCs are 
$2.0 \times 10^6$ yr, $4.6 \times 10^8$ yr, and $8.4 \times 10^8$ yr 
for C1, C2 and C3. 
These results are very different from those of
Scoville et al. (1998), who suggests that  
they may be as old as ($\ga 10^{9}$ years) 
the central globular clusters of  
NGC 5128 (Frogel 1984), because of the similarity of near-infrared colours.
We attribute this slight redness of the star clusters to a little
amount of dust rather than to a very large age.  
We also mention that since the near-infrared colours of star clusters 
mainly depend on the colours of red supergiants and red giants, 
the evolution of near-infrared colours is small for the age 
larger than $\sim 10^7$ yr. 
On the other hand, the optical colours, which mainly determined by 
the colours of the upper main sequence stars, evolve gradually 
from blue to red for the Hubble time. 
It is therefore useful to use the optical colours 
to determine the age of SSCs.

We summarize the best-fitting parameters for the circumnuclear SSCs 
in Table~1. 
The masses of the SSCs 
are about $10^5 {\rm M}_{\odot}$, similar to the 
masses of galactic globular clusters. 
Maybe these are the globular clusters hypothesized to form during mergers
by Zepf \& Ashman (1993), who note that elliptical galaxies (thought to
be merger remnants) have higher specific globular cluster frequencies
than spiral galaxies (thought to be merger progenitors), and that the
metallicity distributions of the globular clusters in the ellipticals
are bimodal.  
The luminosities of the SSCs are about $10^8 {\rm L}_{\odot}$. 
Therefore C1 - 3 contribute negligibly to the total bolometric luminosity
of Arp 220.  This result is unchanged if we consider the other circumnuclear 
SSCs which are detected only at near-infrared wavelengths. 

%----------------------------------------------------------------------------------

\vspace {0.5cm}

We would like to thank Bob Joseph for his useful comments and 
suggestions on an earlier version of our manuscript.
YS thanks the Japan Society for Promotion of Science (JSPS)
Research Fellowships for Young Scientist. 
NT thanks the PPARC for financial support.
This work was financially supported in part by Grant-in-Aids for the Scientific
Research (Nos.\ 07044054, 10044052, and 10304013) of the Japanese Ministry of
Education, Science, Sports and Culture.

%----------------------------------------------------------------------------------
%\newpage

\clearpage
%%%%%%%%%% FIGURE CAPTIONS %%%%%%%%%%%%%%%%%%%%%%
\begin{figure}
%\vspace{5cm}
\caption{
Comparison of the positions of SSCs detected in the optical 
(Shaya et al.~1994: filled circles) and near-infrared 
(Scoville et al.~1998; open circles) images.
The left upper panel shows the positions taken directly from both of these 
papers and the left, lower panel is a close up of the central region.
The contours show the radio continuum image at 4.83 GHz taken from
Baan \& Haschick (1995).
In the right, upper panel, the near-infrared 
image is rotated by 4 degree clockwise as described in the text. 
The right, lower panel 
again shows a close up of the central region.}
\label{fig-1}
\end{figure}

\begin{figure}
\caption{
Comparisons of the observed optical and near-infrared SEDs (filled circles)
with those of best-fit models (dotted lines mean ISB model 
and dashed lines mean CSF model; see the text and 
Table~1 for details)
for the six SSCs analyzed in this paper.
The six points on the SED come from broadband {\it VRIJHK} measurements.
The V band points are all upper limits for the nuclear clusters.}
\label{fig-2}
\end{figure}

\begin{figure}
\caption{
Summary of the plausible parameters of SSCs in the centre of Arp 220 
on $A_V$ - age plane. 
The shaded region is that allowed
by SED fitting at the 99 \% confidence level. 
The dashed lines in (a) and (d) 
represent masses of $1 \times 10^9 {\rm M}_{\odot}$, 
as described in the text. 
The region to the left of this line is permitted.  
The dotted lines in (a) - (f) show the epoch when the strength of 
CO index equal to 0.20 mag for CSF model and ISB model. 
The region between these two lines is permitted.}
\label{fig-3}
\end{figure}

\begin{figure}
\caption{
Comparison of the observed near-infrared SED of N4 (filled circles) 
with the best-fit model SED. 
Dashed line is the best-fit CSF model whose age is $1.0 \times 10^7$ yr 
and the value of $A_V$ is 12.2 mag. 
Dotted line is the best-fit ISB model whose age is $1.7 \times 10^7$ yr 
and the value of $A_V$ is 12.0 mag.}
\label{fig-4}
\end{figure}

\clearpage

\begin{table}
\begin{minipage}{70mm}
\caption{Derived physical properties of the SSCs}
\label{tab-1}
\begin{tabular}{@{}lccccccc}
SSC & ID\footnote{Identifications in Shaya et al. (1994).} 
& ID\footnote{Identifications in Scoville et al. (1998).} 
& SED & Age  & $A_V$  & 
$M_*$\footnote{Stellar mass of SSCs.} 
& $L_{\rm bol}$\footnote{Extinction-corrected bolometric luminosity of SSCs} \\
    &                     &          &  type    &  $(10^7$ y) & (mag) & $(M_\odot)$ & $(L_\odot)$ \\
\hline
N1 & 1 & W  & CSF & 10  & 10.8 & $6.6\times 10^8$ & $9.8\times 10^{10}$ \\
N2 & 3 & S  & CSF & 10  &  8.4 & $6.7\times 10^7$ & $1.0\times 10^{10}$  \\
N3 & 2 & NE & CSF & 10  & 11.4 & $2.5\times 10^8$ & $3.6\times 10^{10}$ \\
\hline
N1 & 1 & W  & ISB & 1   &  8.6 & $8.2\times 10^7$ & $2.4\times 10^{10}$ \\
N2 & 3 & S  & ISB & 1.3 &  7.6 & $2.2\times 10^7$ & $4.3\times 10^{9}$  \\
N3 & 2 & NE & ISB & 1.3 & 10.8 & $7.9\times 10^7$ & $1.6\times 10^{10}$ \\
\hline
C1 & 11 & 4 & ISB & 0.08& 0.4 & $1.1\times 10^5$ & $1.7\times 10^{8}$ \\
C2 &  9 & 1 & ISB & 2.8 & 0.4 & $6.9\times 10^5$ & $5.4\times 10^{7}$ \\
C3 & 10 & 2 & ISB & 3.8 & 0   & $3.5\times 10^5$ & $1.9\times 10^{7}$ \\
\hline
\end{tabular}
\end{minipage}
\end{table}

\clearpage
%%%%%%%%%% APPENDIX %%%%%%%%%%%%%%%%%%%%%%%%%%%%%
\appendix
\section{Onion-skin model of extinction}

\noindent
Throughout this work we have used a screen model to quantify extinction
by dust.  This assumes that all of the dust responsible for the extinction 
is distributed within Arp 220 between the SSCs and the observer, along our
line of sight.  This model is inadequate if the SSCs themselves possess 
significant amount of dust, and we investigate this possibility using the
onion-skin model of Surace \& Sanders (1999).  
In that model, to model the effect of mixed stars and dust, 
a distribution of the unreddened luminosity $L_{A_V}$ obscured by a given
line-of-sight extinction $A_V$ is assumed to be a power-law form; 
$L_{A_V} = L_{A_V=0} (A_V+1)^\alpha$. 
The amount of internal extinction from within the SSC is therefore 
parameterised by two numbers: $A_{V,{\rm max}}$, 
which is a maximum value of $A_V$, and $\alpha$, which is 
a parameter modeling the distribution of
dust within the SSCs (the case $\alpha=0$ corresponds to a scenario in which
the dust and stars are well mixed; this is the case that we are worried
about here).
Figure~A1 shows the SEDs of SSCs for some of these models;
the relevant parameters are presented in Table~A1.
Here $A_{V,s}$ is the extinction from a screen component unassociated with
the SSCs that is superimposed on the internal extinction.
 
The most important thing that we infer from the figure
is that the shape of the SED is 
determined primarily by the foreground extinction.
Were extinction from dust within the clusters to be the more important,
we would expect a far higher visible-to-near-infrared
flux ratio than is observed.  
Our use of the screen model in the SED fitting is therefore reasonable.

It is important to note, however, that, although
the shapes of the SEDs are similar (compare the model 4 with the screen model),
the flux of the well-mixed+screen model is smaller than that of
the pure-screen model. 
This is because the stars with a large local $A_V$
do not contribute to the global SED 
of the SSC at visible and near-infrared wavelengths.
Therefore the SSC luminosities evaluated
in the main body of the paper are 
formally lower limits because of this danger of
there existing very heavily obscured regions {\it within} the SSCs.
These very-heavily obscured regions could be
responsible for a substantial part of the bolometric luminosity of the SSCs
or even the whole galaxy.
Operationally,
as far as the spectral energy distributions are concerned, these regions
behave exactly like very-heavily
obscured regions {\it outside} the SSCs, for example at the galaxy
nucleus, as they have no measurable signature at the wavelengths studied
here.

\begin{figure}
\caption{
The effect on the SEDs of dust within the SSCs, computed from the onion-skin
model described in the text.  
The solid line is the SED of an SSC with an age of 80 Myr for ISB model.
The dotted line is the SED corrected using only a screen model
with a line-of-sight extinction $A_{V,s} = 10$. 
The other lines represent the models listed in Table A1. 
}
\label{fig-a1}
\end{figure}

\begin{table}
\begin{minipage}{70mm}
\caption{Parameters of onion skin model}
\label{tab-A1}
\begin{tabular}{@{}lccc}
model & $A_{V,s}$ & $A_{V,{\rm max}}$ & $\alpha$ \\
%\hline
1& 0 & 10 & 0\\
2& 0 & 100& 0\\
3& 5 & 100& 0\\
4& 10& 100& 0\\
%\hline
\end{tabular}
\end{minipage}
\end{table}


\begin{thebibliography}{}
\bibitem{}Armus L., Heckman T. M., Miley G. K. 1989, ApJ, 347, 727
\bibitem{}Armus L., Neugebauer G., Soifer B. T., Matthews K. 
              1995, AJ, 110, 2610
\bibitem{}Ashman K. M., Zepf S. E. 1992, ApJ, 384, 50
\bibitem{}Baan W. A., Haschick A. D. 1995, ApJ, 454, 745
\bibitem{}Cardelli J. A., Clayton G. C., Mathis J. S. 1989, 
              ApJ, 345, 245
\bibitem{}Clements D.~L., Sutherland W.~J., McMahon R.~G., \& Saunders 
              W.~1996, MNRAS, 279, 477
\bibitem{}de Vaucouleurs G., de Vaucouleurs A., Corwin H. G., Jr., Buta R. J.,
              Paturel G., Forqu\'e P. 1991, Third Reference Catalogue of Bright
              Galaxies (Springer-Verlag)
\bibitem{}Downes D., Solomon P. M. 1998, ApJ, 507, 615
\bibitem{}Faber S.~M. 1973, ApJ, 179, 731  
\bibitem{}Genzel R. et al. 1998, ApJ, 498, 579
\bibitem{}Goldader J. D., Joseph R. D., Doyon R., Sanders, D. B.
              1995, ApJ, 444, 97
\bibitem{}Goldader J. D., Joseph R. D., Doyon R., Sanders D. B.  
              1997, ApJS, 108, 449
\bibitem{}Holtzman J. A., et al. 1992, AJ, 103, 691
\bibitem{}Kim D.-C., Sanders D. B., Veilleux S., Mazzarella J. M., 
              Soifer B. T. 1995, ApJS, 98, 171
\bibitem{}Larkin J. E., Armus L., Knop K., Matthews K., Soifer, B. T.
              1995, ApJ, 452, 599
\bibitem{}Leitherer C. et al. 1999, ApJS, 123, 3
\bibitem{}Lutz D. 1991, A \& A, 245, 31
\bibitem{}Mouri H., Taniguchi Y. 1992, ApJ, 386, 68
\bibitem{}Murphy T.~W., Armus L., Matthews K., Soifer B.~T., 
      Mazzarella J.~M.~1996, AJ, 111, 1025
\bibitem{}Prestwich A. H., 
              Joseph R. D., Wright G. S. 1994, ApJ, 422, 73
\bibitem{}Rieke G. H., Cutri R. M., Black J. H., Kailey W. F., McAlary C. W., 
Lebofsky M. J., Elston R. 1985, ApJ, 290, 116
\bibitem{}Rieke G. H., Lebofsky M. J. 1985, ApJ, 288, 618
\bibitem{}Sakamoto K.. Scoville N. Z., Yun M. S., Crosas M., Genzel R.,
              Tacconi L. J. 1999, ApJ, 514, 77
\bibitem{}Sanders D. B., Mirabel I. F. 1996, ARA \& A, 34, 749
\bibitem{}Sanders D. B., Soifer B.~T., Elias J.~H., 
Madore B.~F., Matthews K., Neugebauer G., Scoville N.~Z.~1988,
ApJ, 325, 74
\bibitem{}Savage B. D., Mathis J. S. 1979, ARA\&A, 17, 73
\bibitem{}Scoville N. Z. et al. 1998, ApJ, 492, L107
\bibitem{}Scoville N. Z., Sargent A. I., Sanders D. B., Soifer B. T.
              1991, ApJ, 366, L5
\bibitem{}Scoville N. Z., Yun M. S., Bryant P. M. 1997, ApJ, 484, 702
\bibitem{}Shaya E., Dowling D. M., Currie D. G.,  Faber S. M.,  
              Groth E. J. 1994, AJ, 107, 1675
\bibitem{}Soifer B. T., Neugebauer G., Matthews K., Becklin E.~E.,
Ressler M., Werner M.~W., Weinberger A.~J., Egami E.~1999, ApJ, 513, 207
\bibitem{}Surace J. A., Sanders D. B. 1999, ApJ, 512, 162
\bibitem{}Surace J. A., Sanders D. B., Vacca W. D., Veilleux S., 
              Mazzarella J. M. 1998, ApJ, 492, 116
\bibitem{}Zepf S. E., Ashman K. M. 1993, MNRAS, 264, 611
\end{thebibliography}
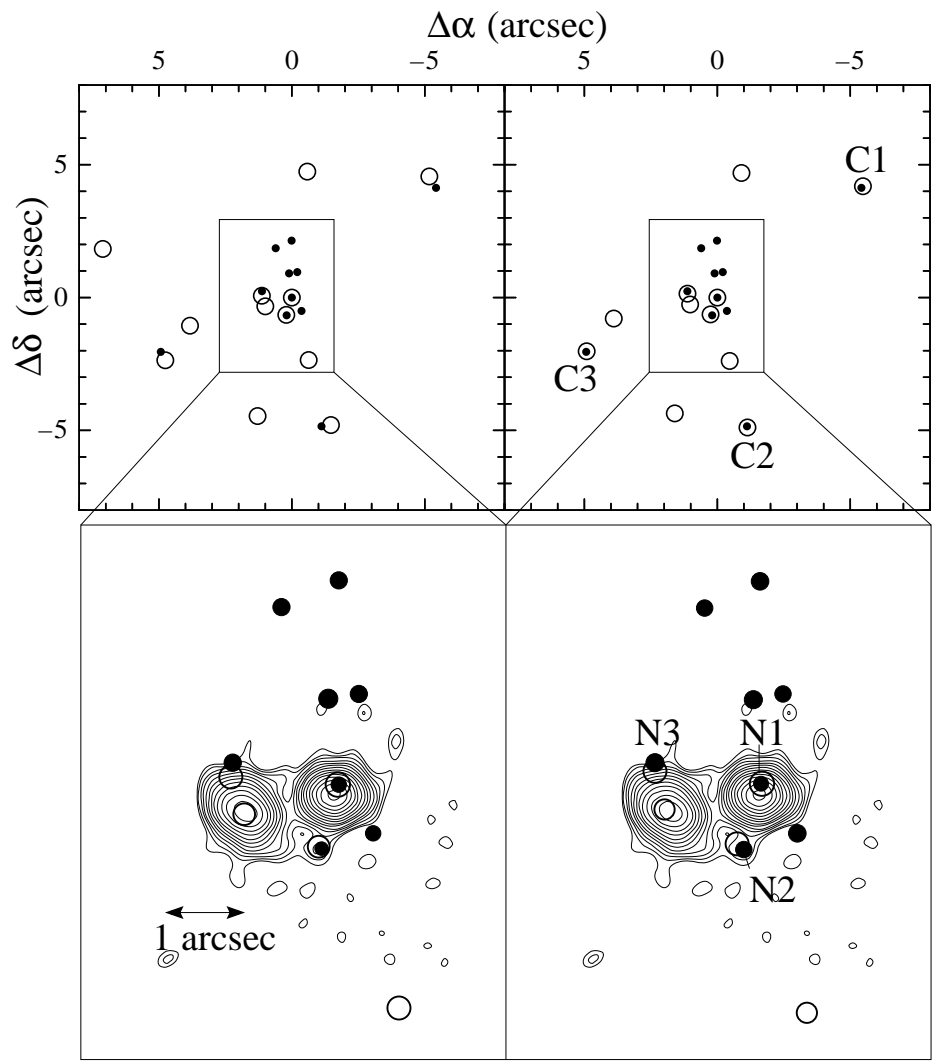
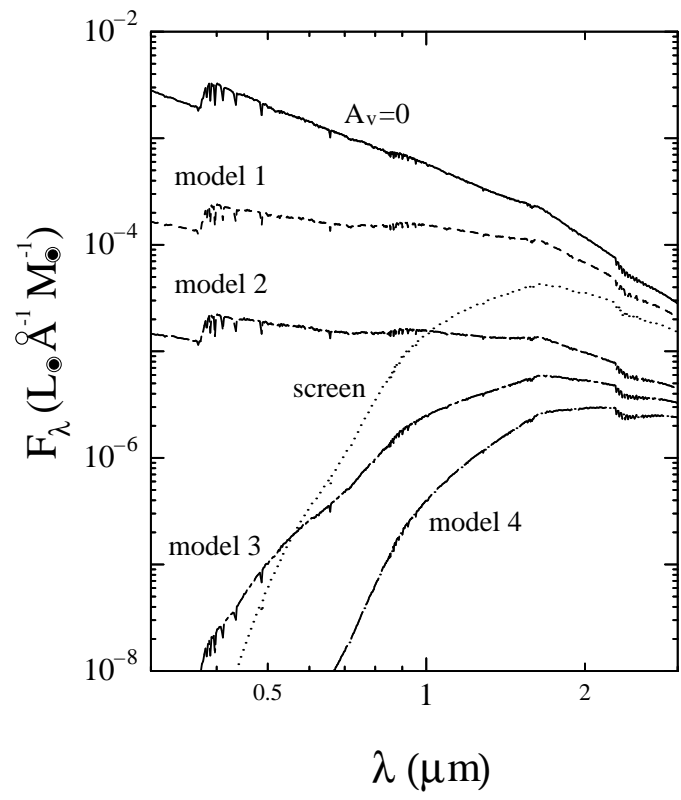
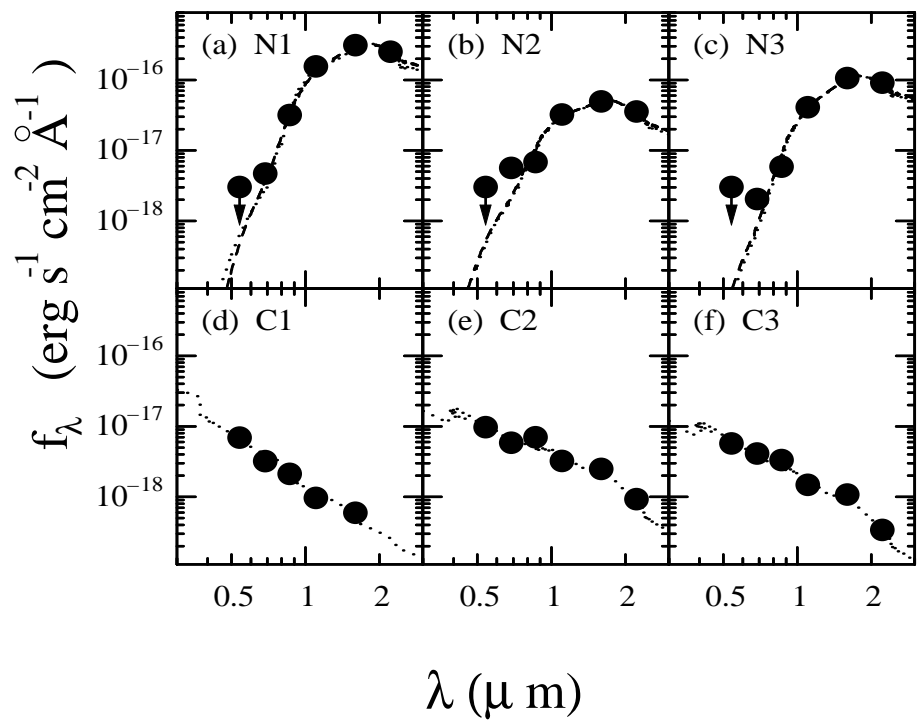
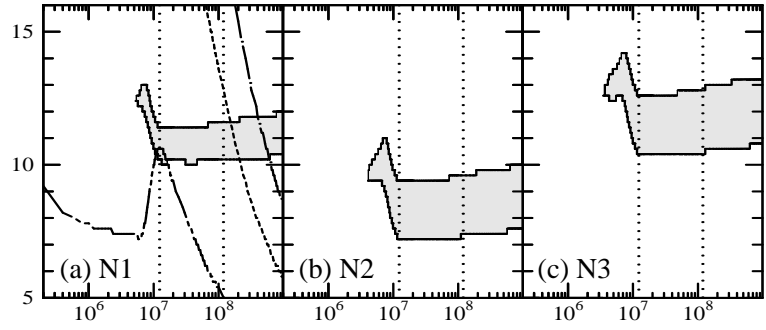
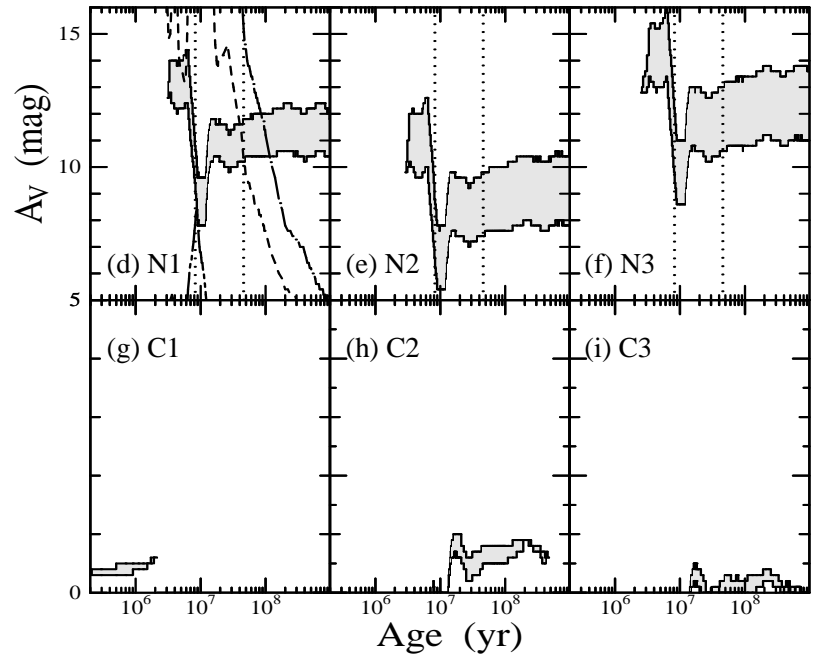
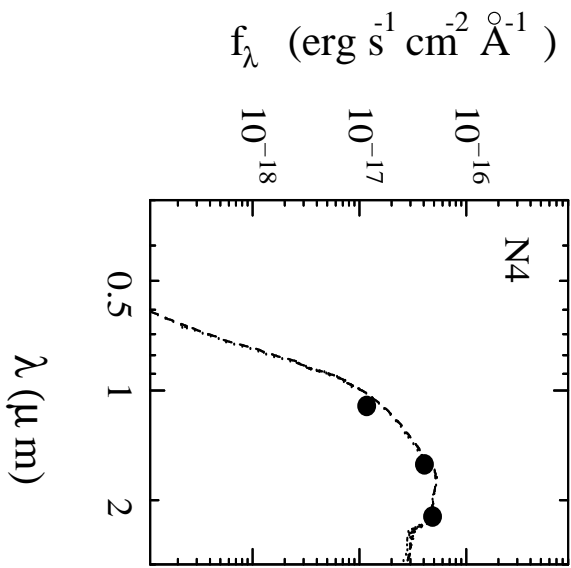
\end{document}